# Tube into pearls: A membrane-driven pearling instability shapes platelet biogenesis

C. Léon[1]*, N. Brassard-Jollive[1], D. Gonzalez-Rodriguez[2], D. Riveline[3,4,5,6,7]*

[1]Université de Strasbourg, INSERM, EFS Grand Est, BPPS UMR-S 1255 ; Strasbourg, France
[2] Université de Lorraine, LCP-A2MC, F-57000 Metz, France
[3]Laboratory of Cell Physics ISIS/IGBMC, CNRS and University of Strasbourg; Strasbourg, France
[4]Institut de Génétique et de Biologie Moléculaire et Cellulaire ; Illkirch, France
[5]Centre National de la Recherche Scientifique, UMR7104 ; Illkirch, France
[6]Institut National de la Santé et de la Recherche Médicale, U964 ; Illkirch, France
[7]Université de Strasbourg ; Illkirch, France

*Corresponding authors.
Email: catherine.leon@efs.sante.fr- riveline@unistra.fr

**Running title:** Blood platelet formation as a physiological example of a pearling instability

**Abstract**

At the end of the 19th century, Rayleigh and Plateau explained the physical principle behind the fragmentation of a liquid jet into regular droplets commonly observed in everyday life from a faucet. The classical Rayleigh-Plateau instability concerns liquid jets governed by inertia and surface tension, whereas biological tubes are membrane-bounded and inertia-free. We therefore refer to the process observed here as a pearling instability, formally analogous to Rayleigh-Plateau but dominated by membrane mechanics. Although pearling-type instabilities have long been recognised in lipid tubes and some biological systems, a clear physiological example remained elusive. Here, we present results showing that pearling instability occurs during the physiological process of platelet formation. Platelets are formed from megakaryocytes in the bone marrow by the extension of long protrusions, called proplatelets, that traverse the blood vessels. As they extend in the bloodstream, proplatelets become pearled and detach. Long and pearled proplatelets then circulate in the peripheral blood before their fragmentation into calibrated platelets. We propose that this pearling, by creating regular constrictions along the proplatelet, is key to the process of proplatelet fragmentation into individual platelets of calibrated size. Pearling instability thus acts as a mechanobiological regulator allowing local delivery of the right size platelets to the right place at the right time. Our observations quantitatively match parameter-free theoretical predictions for membrane pearling, supporting a unified physical picture.

**Rayleigh-Plateau and pearling instabilities: transformation of tubes into pearls below critical dimensions.**

The pearling instability has attracted significant attention due to its substantial relevance in various industrial processes, such as inkjet printing technologies or single cell encapsulation among many others. This well-documented phenomenon occurs when a slender liquid jet or thread undergoes fragmentation into droplets (Fig. 1A) (*1*). This transformation is primarily governed by instabilities and perturbations within the jet, as initially highlighted by Savart in the mid-19$^{th}$ century (*2*) and treated 50 years later by Plateau (*3*). Subsequently, Rayleigh introduced a theoretical framework (*4*). Surface tension is the driving force behind this instability primarily driven by the minimization of surface energy.

The Rayleigh-Plateau instability of a liquid cylinder emerges from the competition of inertia and surface tension. It can be summarized by the following statement: a cylindrical column of liquid becomes unstable when its length exceeds π times its diameter (critical diameter)(*3, 4*) (Fig. 1A). However, in biological contexts, inertia is negligible making Rayleigh's original solution inapplicable. Indeed, biological systems such as lipid tubes (*5*), mitochondria (*6*), or cells consist of a viscous fluid enclosed by a deformable membrane rather than a fluid-fluid interface. This alters the boundary conditions and resulting dynamics. We therefore adopt the broader term *pearling instability* for describing the dynamics of membrane-bounded tubes. Remarkably, despite distinct underlying physical mechanisms, the pearling instability emerges at a characteristic wavelength-to-diameter ratio close to that predicted by the Rayleigh-Plateau instability, though with a slight deviation. Since its description, the pearling instability has attracted much attention from the scientists who evaluated the influence of physical alterations of the fluids in a variety of situations with different physical mechanisms including changes in viscosity, the presence of shear forces, elasticity of the bulk phase or surfactant concentration (*1*). The interplay between capillary and elastic stresses leads to the formation of thin and stable filaments between drops, akin to a 'pearls-on-a-string' structure, followed by various generation of smaller drops forming amidst the initially formed pearls that adds complexity to the phenomenon (*1*).

**Distinct physical mechanisms in membranes and in cells.**

While the core concept of pearling resulting from the Rayleigh-Plateau instability was extensively explored from a physical point of view, its significance in the field of biology was later on highly questioned. Several studies thoroughly examined pearling instability of lipid vesicles using experimental or theoretical approaches (*5*). A pattern very similar to the "pearl-on-a-string" was observed on cylindrical vesicles enclosing Newtonian fluids (*5*). This pearling instability can be induced through various methods. Pulling membrane tethers with optically trapped particles(*7, 8*), protein mediated anchoring of membrane tethers to a substrate (*9*), applying a magnetic field (*10*), electric field (*11*) or osmotic pressure gradient (*12, 13*) can all lead to pearling. Considering cell membranes adds an extra layer of complexity due to the presence of bending and cytoskeletal proteins that span or interact with the lipids to control cortical tension and curvature. Modulation of the cytoskeleton network not only helps cells to constantly adapt to various environmental factors but has also been shown to control cell pearling. Its disruption using actin depolymerizing drugs leads adherent cells to alter their shape and ultimately form pearls connected by slender membrane tubes (*9, 14*). Similarly, *in vitro* axons where microtubules or the coupling between F-actin and microtubules is destabilized become susceptible to pearling instability (*15*). This phenomenon could provide an explanation for the pearled appearance of axons before they undergo atrophy in various neurodegenerative conditions (*15*). Therefore, the fundamental physics associated with the pearling instability applies to living cells both *in vitro* and *in vivo* under pathological degenerative conditions. However, a genuine biological role for this phenomenon was still missing. Physiological applications of such instabilities have more recently been explored, from a theoretical standpoint in tissue growth (*16*) and through a combination of theory and experiments in mitochondrial shape regulation (*6*). Using theoretical calculations and computer simulations, Bächer *et al*. proposed a role for a pearling instability in flow-accelerated platelet biogenesis (*17*). In this commentary, we present biological observations that pearling instability is indeed physiologically relevant *in vivo* when considering the specialized mechanism of platelet formation.

**Proplatelet formation: a force balance mechanism with pearling instabilities.**

Each day, the human body generates approximately $10^{11}$ blood platelets to ensure the maintenance of hemostasis and prevent bleeding. Platelets are small, anucleate and calibrated discoid fragments (1.5-2 and 2.5-3 µm in diameter for mouse and human platelet, respectively) produced by their parent cells called megakaryocytes (MKs). MK themselves differentiate from hematopoietic stem cells in the bone marrow, where they undergo a switch from mitosis to endomitosis resulting in polyploid giant cells (*18*). Platelets are produced by unique mechanisms: once an initial MK protrusion has pierced through the sinusoid vessels to reach the fluid blood, buds and longer protrusions (the proplatelets) are formed (*18, 19*). The proplatelets, still attached to the MK body by one end, extend in the bloodstream. This is made possible by the presence of a large intracellular membrane reservoir, known as the demarcation membrane system (DMS), which fuses with the plasma membrane to fuel in lipids and membrane components (Fig. 2). Once detached from its mother cell, the proplatelet will fragment during successive passages through the lung microcirculation to release individual platelets of homogeneous size (*18, 20*).

Real time *in vivo* observation of proplatelet elongation is possible by visualizing bone marrow inside the mouse skull bone using two-photon microscopy (Fig. 1B-Ci) (*19*). Most of the proplatelets, as they elongate, present undulations along their cylindrical shape that can transition into pearls, very reminiscent of the pearl-on-string configuration observed with membrane tubes. The pearls have an average minor axis diameter of 6.8 ±2.5 µm (major axis is 17.1±6.4, mean±SD). At that stage, pearling is unstable as pearled and unpearled states fluctuate along proplatelets, with relative heterogeneity in their diameter and inter-pearl distance (Fig. 1B ,Ci, Civ). Once released in the bloodstream, the pearled proplatelet fragments transiently circulate in the peripheral blood and mature to adopt a more homogeneous pearl profile (mean±SD of pearls from human circulating proplatelets are 1.8±0.6 and 3.5±1.0 µm for minor and major diameter of pearls, respectively) (Fig. 1Cii, Civ).

*In vivo*, hemodynamics significantly contribute to increase proplatelet elongation velocity (*17, 19*). However, it is not a prerequisite for proplatelet pearling. Under static *in vitro* culture, proplatelets also exhibit the pearl-on-string configuration (Fig. 1Ciii, Civ), with pearl diameter (D) and wavelength (λ) being more homogeneously distributed compared to *in vivo* (Fig. 1Civ). Interestingly, although absolute pearl size differs between conditions, the relative heterogeneity, measured as the standard deviation divided by the mean, remains approximately constant and about 100%. The robust emergence of regularly spaced pearls across all conditions supports the idea that a common physical mechanism consistent with a pearling instability is involved during proplatelet elongation.

To test this hypothesis, we measured D and λ of proplatelet fragments *in vivo* (mouse), *ex vivo* (human blood), and *in vitro* cultures (human). These measurements are compared to the theoretical prediction for the pearling instability, as derived in the literature (*6*) (Fig. 2A) and λ= 4.25 D where D and λ correspond to the main diameter of the pearl and to the distance between pearls respectively. The match is striking and fully quantitative as the theory involves no adjustable parameters. The theoretical model predicts a characteristic ratio between D and λ, corresponding to the fastest-growing mode of a viscous liquid cylinder enclosed by a membrane. This aspect ratio is well conserved across conditions and matches values observed in other biological systems (*6*). This further confirms that the pearling instability constitutes the fundamental physical mechanism behind proplatelet fragmentation.

While this sets a baseline, the aspect ratio can be slightly modulated by additional factors such as external shear flow, which tends to produce more elongated pearls (*in vivo* and *ex vivo* conditions in Fig. 2A). However, the most significant variability concerns pearl size (D), which changes markedly both across and within conditions. This variability likely reflects a combination of biological and physical influences, including membrane supply, cytoskeletal tension, and flow conditions. A full understanding of pearl size requires considering the interplay between two dynamical processes: the development of the pearling instability and the elongation of the proplatelet. Elongation results from a balance of forces, including external shear, membrane reservoir dynamics, and contractility, that together regulate membrane tension and geometry (Fig. 2B). This force balance governs not only the occurrence of pearling but also its precise characteristics, ultimately shaping platelet formation.

Overall, this underscores the significance of the pearling phenomenon that is intrinsic to the process of platelet biogenesis whatever the extracellular environment (*in vitro* culture or bone marrow) or the proplatelet formation stage (MK-attached proplatelet or circulating proplatelet). Through the formation of pearls, pearling contributes to the homogeneous size of the final platelet product, a prerequisite for their normal function. The presence of abnormally large or small platelets, mostly resulting from cytoskeletal disorder, is associated with abnormal platelet function and impaired hemostasis and thrombosis (*18*).

The greater pearl size in proplatelets attached to MKs observed *in vivo* in the bone marrow compared to *in vitro* conditions (Fig. 1B, Ci,iv) is likely due to differences in the molecular and biophysical mechanisms involved in proplatelet formation. *In vitro*, microtubules are required for elongation while they are dispensable *in vivo* (*19*). Indeed, *in vivo* external shear flow raises membrane tension and extends proplatelets, while actomyosin and microtubules modulate tension locally. Both parameters enter the dynamics of propatelet elongation thereby tuning pearl shape. Furthermore, actomyosin, by imparting cortical tension, tends to counteract the elongating flow forces, resulting in occasional pauses and, at times, even transient slight retraction of the proplatelets (*19*). The precise regulation of the balance between protrusive and retraction forces is not fully understood, although it can be hypothesized that shear forces may play a role in enhancing actomyosin activity to resist flow forces. In turn, the actomyosin, by controlling internal tension, most probably conditions the thinning of the tube, its membrane deformability, hence its undulations, counterbalanced by the intraproplatelet hydrostatic pressure.

Hence, the balance between longitudinal dynamics and transversal dynamics will condition the formation of pearls (*21*). *In vivo*, although not *in vitro*, absence of myosin indeed results in proplatelets that are devoid of pearls, thinner and more rapidly elongating *(19). In vitro,* defects in actin dynamics prevent pearling (*22*). Future experiments, such as microfluidic assays with controlled flow environments, *in vivo* bone marrow flow modulation following topical application of pharmacological drugs on the skull bone (*23*) or pharmacological or genetic modulation of actomyosin contractility, could help quantify the respective contributions of external shear and active tension generation to the pearling instability. These experiments could be associated to the measurements of each force component to test quantitatively each term and establish connections between these specific molecular mechanisms and the mesoscopic shapes of pearling instability.

Another dimension to consider is the role played by the internal DMS membrane reservoir. The DMS fuels the plasma membrane of the elongating proplatelet by fusion all along the proplatelet under spatio-temporal regulations that are unknown. Hence, each fusion temporarily and substantially releases membrane tension until further elongation. A balance between cellular membrane thinning and acquisition of membrane reservoir could determine the overall dynamics. Once detached from its mother cell, the circulating proplatelet adopt homogeneous platelet-sized pearls.

An unresolved and significant question revolves around the mechanisms underlying the detachment of proplatelets from their mother cells and ultimately the release of proper platelets. Release sites are zones of pinching between pearls (*19, 24*). One hypothesis is that similar to the cell division cleavage furrow that pinches the cytoplasm into two lobes, the pearling instability promotes constrictions that help in the normal detachment of these fragments. The involvement of an active machinery as that reported for membrane fission (*25*) is still under investigation.

Therefore, although different from the physical phenomenon involving liquid jet thinning as initially described by Rayleigh and Plateau 150 years ago, the pearling instability appears to be also of central biological significance. A more thorough exploration of its exact role in proplatelet and platelet formation will be important. An essential aspect to consider involves discerning the respective impact of pure passive (mechanical) and active (mechanobiological) responses in these processes, along with their potential interaction. Furthermore, it is crucial to assess the influence of flow and the various cytoskeletal or bending proteins in the development of pearls and constrictions that condition the final platelet size. Exploring the possible interplay between physical and biological mechanisms is especially important to offer novel approaches to understand hemostasis defects and optimize *in vitro* platelet production.

**Acknowledgments:** The authors wish to thanks Alicia Bornert and Josiane Weber for two-photon and mono-photon images and Andrei K Garzon Dasgupta (inserm U1255-EFS Grand Est) for critical reading of the manuscript. We also thank Christophe Clanet (Ladhyx, Ecole Polytechnique, France) for providing F. Savart's scheme. We thank ANR (Agence Nationale de la Recherche) for funding (grant ANR-18-CE14-0037-01 and ANR-23-CE52-0012. N. B-J. was supported by ARMESA (Association de Recherche et Développement en Médecine et Santé Publique).

**Legend to Figures**.

**Figure. 1. Pearling phenomenon behind platelet formation;** A) Scheme adapted from Felix Savart (1833), illustrating pearling formation following liquid jet thinning (*2*). B) Upper, scheme of the proplatelet (white arrow) initiation, elongation and release occurring in the bone marrow sinusoid vessels (*sv*). Lower, time lapse showing *in vivo* mouse proplatelet elongation (green, Green Florescent Protein (GFP) labeling) and rupture; z-stack with maximal projection, time is minute; scissors denote the site of rupture; red is the microvessel fluorescent tracer; scale bar is 20 µm; from (*19*) with permission; *, MK, located above the vessel; $\vec{F}_{flow}$, force from blood flow. C) i) *In vivo* observation of an elongating proplatelet inside sinusoid vessel (*sv*), attached to its MK body located above the vessel (*) (green, GFP). Observation in the mouse skull bone marrow (z-stack with maximal projection); red is the microvessel fluorescent tracer; white arrow denotes flow direction; from (*19*) with permission. ii) Confocal image showing circulating proplatelet in human peripheral blood (arrow) observed among discoid platelets; immunolabeling with anti-α-tubulin (green) and anti-GPIbβ (red, (pro)platelet membrane marker). iii) Bright field image showing *in vitro* proplatelet obtained from human peripheral blood CD34[+]-derived MKs (as in (*24*) but kept in static culture condition). iv) Scheme for the quantification of pearling followed by (top) measurement of the pearl minor diameter (D) (n=34 pearls from *in vivo* mouse proplatelets, 101 pearls from circulating human proplatelets, 457 pearls from *in vitro* human proplatelet) (of note, major diameter measurement give similar distributions); (bottom), measurement of the distance between adjacent pearls (wavelength, λ) (n=24 for *in vivo* mouse proplatelets, n=76 for circulating human proplatelets, n=373 for *in vitro* human proplatelets).

**Figure 2. Physical mechanisms associated to proplatelet formation and pearling.** A) Wavelength (λ) *vs.* major diameter (D) curve for *in vivo* (green), *ex vivo* (red) and *in vitro* (grey) proplatelet pearls. The dashed line is the theoretical prediction λ= 4.25 D. B) Simplified force balance equation along elongation axis

associated to proplatelet formation and pearling. Once in the blood flow, hemodynamic flow forces ($\vec{F}_{flow}$) and forces resulting from the fusion of the DMS membrane reservoir with the plasma membrane ($\vec{F}_{reservoir}$) contribute to proplatelet elongation, while actomyosin contractile forces ($\vec{F}_{contractility}$) and internal friction forces arising from proplatelet viscosity ($\vec{F}_{friction}$) tend to oppose elongation. More complex force balance equations could be associated to additional terms modulating the differences in longitudinal and transversal velocities of the tube. Microtubules are represented (green) although *in vivo* their role in elongation *per se* is minor. Dc corresponds to a critical diameter for the reversible onset of the stability considering a maximal length reachable by the propatelet.

# Figure 1: Pearling phenomenon behind platelet formation

**A** Rayleigh-Plateau instability

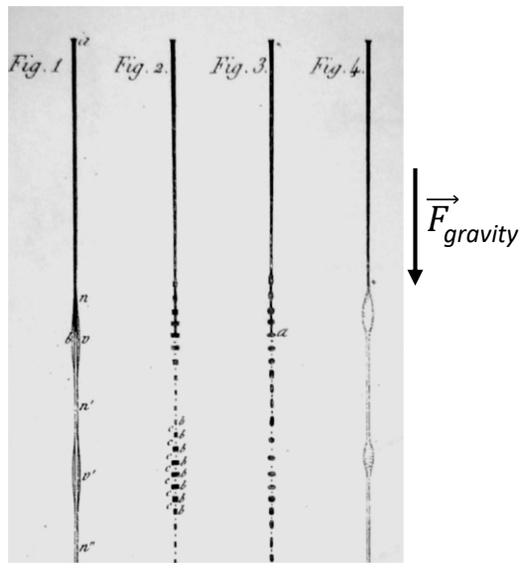

**B** initiation — elongation — release

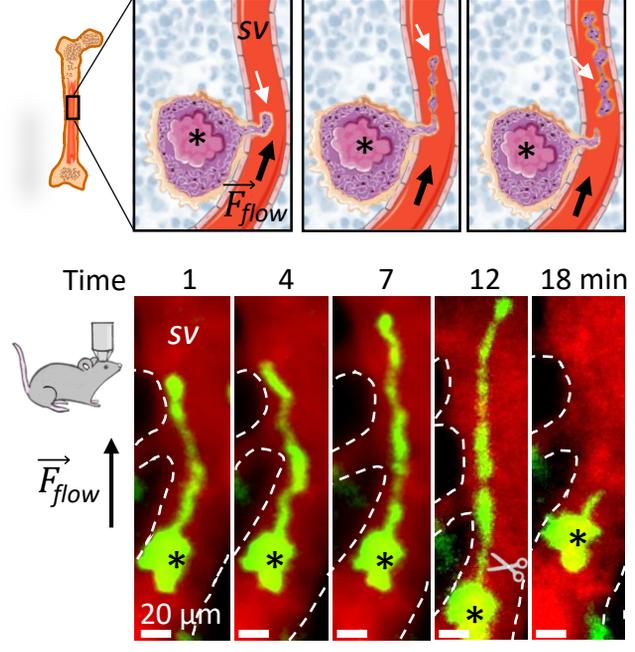

**C**

i  ii  iii  iv Wavelength (λ), Diameter (D)

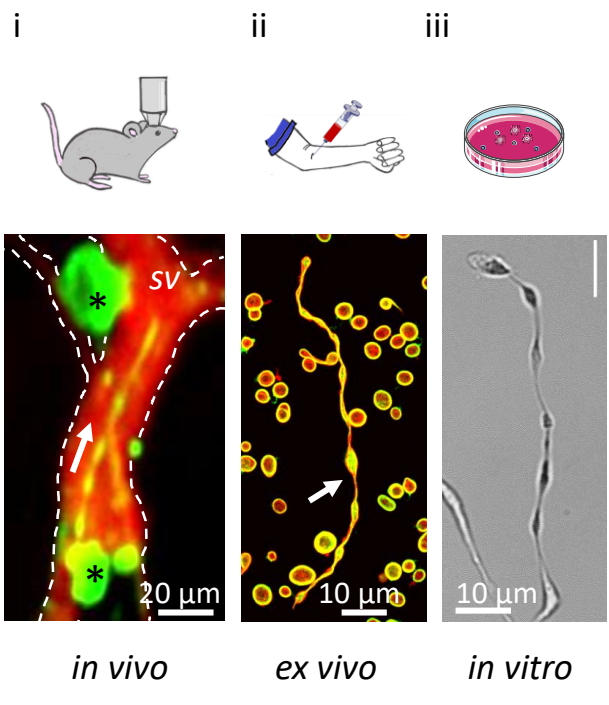

in vivo — ex vivo — in vitro

# Figure 2: Physical mechanisms associated to proplatelet formation and pearling

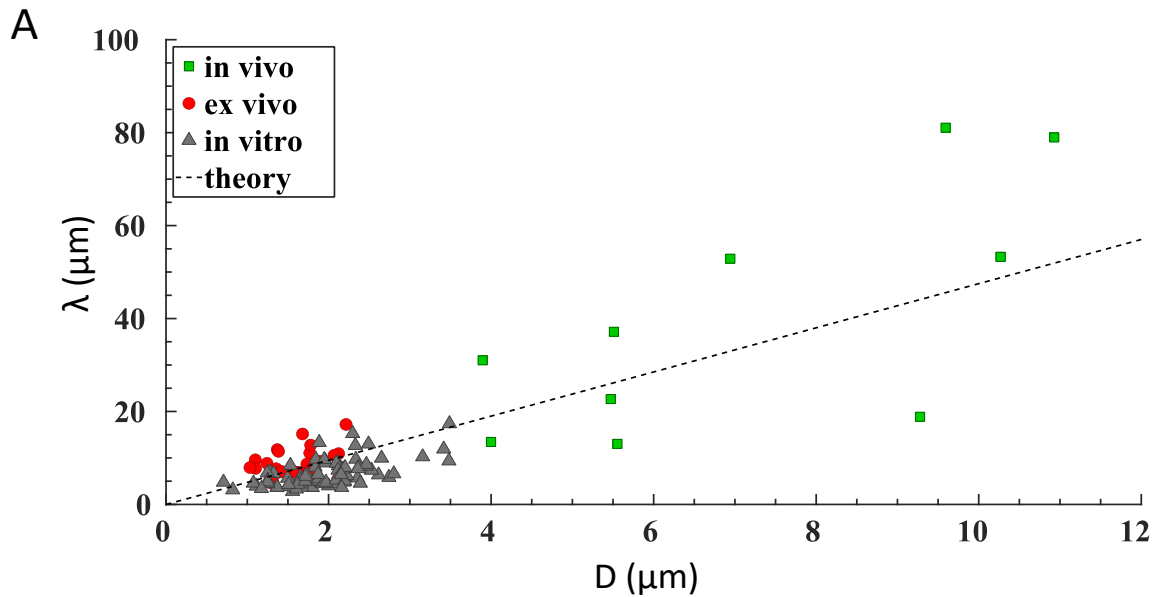

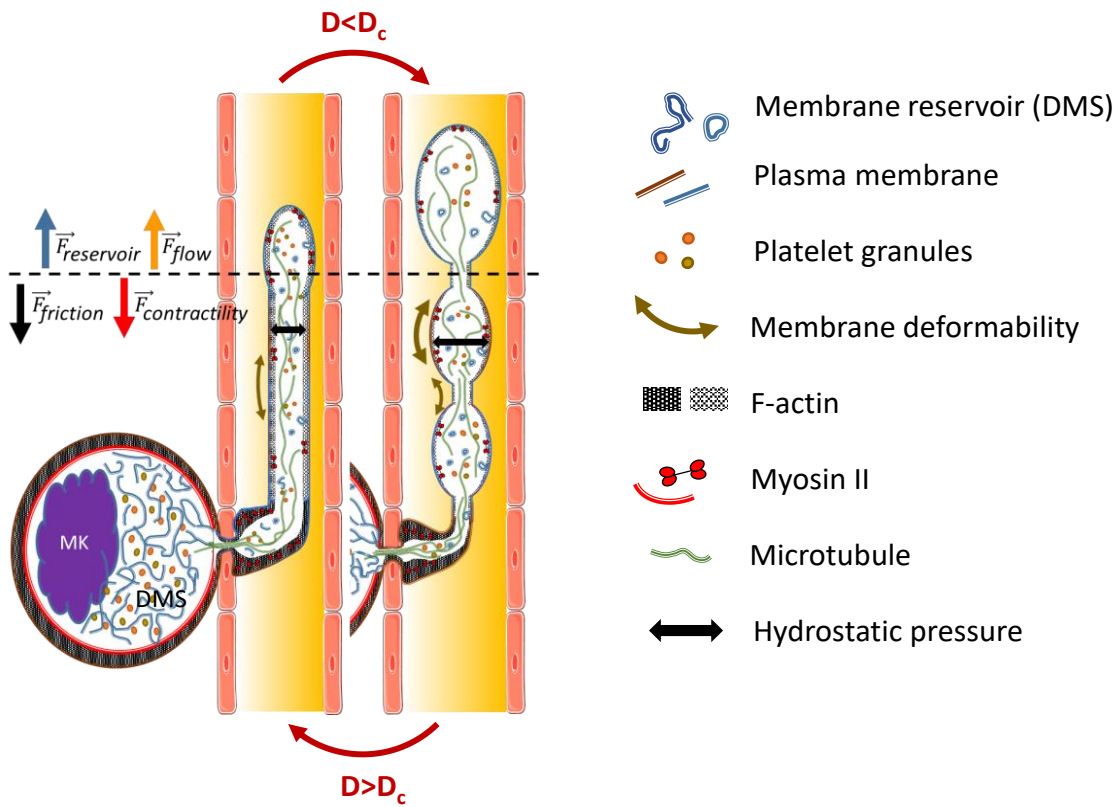